\documentclass[sigconf,9pt]{acmart}

\usepackage[english]{babel}
\usepackage{blindtext}

\usepackage{tikz}
\usetikzlibrary{positioning,arrows.meta, arrows.meta, calc, decorations.markings, matrix,shapes.misc, fit, backgrounds}
\usetikzlibrary{external}
\usepackage{tikzscale}
\usepackage{adjustbox}
\usepackage{amsmath}
\usepackage{algorithm}
\usepackage{subcaption}
\usepackage{csquotes}
\usepackage{dirtytalk}
\usepackage{xspace,setspace}
\usepackage{listings}
\usepackage{makecell}
\usepackage{epigraph}
\usepackage[compact]{titlesec}
\usepackage[noend]{algpseudocode}

\usepackage{booktabs,calc,lipsum}
\usepackage{colortbl}
\usepackage{wasysym}
\usepackage{booktabs}
\usepackage[inline]{enumitem}
\usepackage{amsthm}
\usepackage{thmtools}
\usepackage{thm-restate} 
\usepackage{apptools}
\usepackage{multirow}
\usepackage[normalem]{ulem}
\usepackage{pifont,stmaryrd}
\usepackage{pgfplots}
\usepackage[framemethod=TikZ]{mdframed}
\pgfplotsset{width=\columnwidth,compat=1.9}

\usepackage[]{changes}

\definecolor{wildstrawberry}{rgb}{0.858, 0.188, 0.478}

\definechangesauthor[color=blue]{George}
\definechangesauthor[color=blue]{Todd}
\definechangesauthor[color=wildstrawberry]{Ryan}
\definechangesauthor[color=orange]{Alan}

\newcommand{\inliststyle}[1]{\textbf{\smaller#1}}
\newlist{inlist}{enumerate*}{1}
\setlist[inlist]{label={\inliststyle{(\arabic*)}}}

\usepackage{ifthen}

\newboolean{TodoNotes}
\newcommand{\IfTodoNotesElse}[2]{\ifthenelse{\boolean{TodoNotes}}{#1}{#2}}

\newboolean{UnivNetwork}
\newcommand{\UniversityNetworkInfo}[1]{\ifthenelse{\boolean{UnivNetwork}}{#1}{\empty}}

\setboolean{UnivNetwork}{False}
\setboolean{TodoNotes}{True}

\definecolor{wildstrawberry}{rgb}{0.858, 0.188, 0.478}
\definecolor{lightgreen}{RGB}{234,249,240}
\definecolor{lightred}{RGB}{255,239,243}
\definecolor{lightblue}{RGB}{220,230,255}

\IfTodoNotesElse{
\def\tdm#1{\textcolor{blue}{\textsf{TDM: #1}}}
\def\ky#1{\textcolor{red}{\textsf{KJ: #1}}}
\def\rb#1{\textcolor{wildstrawberry}{\textsf{RB: #1}}}
\def\at#1{\textcolor{orange}{\textsf{AT:#1}}}
\def\sk#1{\textcolor{brown}{\textsf{SK:#1}}}
\def\gv#1{\textcolor{brown}{\textsf{GV: #1}}}
\def\ez#1{\textcolor{purple}{\textsf{EZ: #1}}}
}
{
\def\tdm#1{}
\def\ky#1{}
\def\rb#1{}
\def\at#1{}
\def\sk#1{}
\def\gv#1{}
\def\ez#1{}
}

\setcopyright{acmcopyright}

\newcommand{\tool}{\textsc{Lightyear}}

\newcommand{\fromisp}{\texttt{FromISP1}\xspace}
\newcommand{\waypoint}{\texttt{WaypointR}\xspace}
\newcommand{\import}{\texttt{Import}\xspace}
\newcommand{\export}{\texttt{Export}\xspace}
\newcommand{\originate}[0]{\texttt{Originate}\xspace}
\newcommand{\recv}[3]{\texttt{recv}(#2 \rightarrow #1, #3)}
\newcommand{\slct}[2]{\texttt{slct}(#1, #2)}
\newcommand{\frwd}[3]{\texttt{frwd}(#1 \rightarrow #2, #3)}

\newcommand{\comm}[1]{\textrm{Comm}(#1)}

\newcommand{\reject}{\textsc{Reject}\xspace}
\newcommand{\traces}{\textsc{Traces}\xspace}
\newcommand{\valid}{\textsc{Valid}\xspace}
\newcommand{\routes}{\textsc{Routes}\xspace}
\newcommand{\routers}{\textsc{Routers}\xspace}
\newcommand{\externals}{\textsc{Externals}\xspace}
\newcommand{\edges}{\textsc{Edges}\xspace}

\setlist{nolistsep}

\copyrightyear{2023}
\acmYear{2023}
\setcopyright{rightsretained}
\acmConference[ACM SIGCOMM '23]{ACM SIGCOMM 2023 Conference}{September 10, 2023}{New York, NY, USA}
\acmBooktitle{ACM SIGCOMM 2023 Conference (ACM SIGCOMM '23), September 10, 2023, New York, NY, USA}
\acmDOI{10.1145/3603269.3604842}
\acmISBN{979-8-4007-0236-5/23/09}

\begin{document}
\title{\tool: Using Modularity to Scale BGP Control Plane Verification}


\author{Alan Tang}
\affiliation{%
  \institution{UCLA}
}
\author{Ryan Beckett}
\affiliation{%
  \institution{Microsoft}
}
\author{Steven Benaloh}
\affiliation{%
  \institution{Microsoft}
}
\author{Karthick Jayaraman}
\affiliation{%
  \institution{Microsoft}
}
\author{Tejas Patil}
\affiliation{%
  \institution{Microsoft}
}
\author{Todd Millstein}
\affiliation{%
  \institution{UCLA}
}
\author{George Varghese}
\affiliation{%
  \institution{UCLA}
}

\renewcommand{\shortauthors}{Tang et al.}

\begin{abstract}
Current network control plane verification tools cannot scale to large networks because of the complexity of jointly reasoning about the behaviors of all network nodes.  We present a \emph{modular} approach to control plane verification, where end-to-end network properties are verified via a set of purely \emph{local} checks on individual nodes and edges.  The approach targets verification of reachability properties for BGP configurations, and provides guarantees in the face of arbitrary external route announcements 
and, for some properties, arbitrary node/link failures. We have proven the approach correct and implemented it in a tool \tool.  Experimentally we show \tool\ scales dramatically better than prior control plane verifiers.  Further, \tool\ has been used for six months to verify properties of a major cloud provider network containing hundreds of routers and tens of thousands of edges, finding and fixing bugs in the process.  To our knowledge no prior control-plane verification tool has been shown to scale to that size and complexity.  
Our modular approach also makes it easy to localize 
configuration errors and enables incremental re-verification.
\end{abstract}



\maketitle

\addtolength\abovedisplayskip{-0.5\baselineskip}%
\addtolength\belowdisplayskip{-0.5\baselineskip}%

\section{Introduction}

Routing in networks today is controlled using low-level configuration on individual routers, which often leads to errors, potentially causing a network outage. 
Many earlier techniques try to remedy this by verifying configurations against specified end-to-end network behavior. For instance, Minesweeper~\cite{minesweeper} models network behavior using SMT constraints, ARC~\cite{ARC} and Tiramisu~\cite{tiramisu} use graphs, and Plankton~\cite{plankton} uses explicit-state model checking. 

These techniques provide strong guarantees, frequently reasoning about network behavior over all possible external announcements and/or link failures.  However, a key open problem is to scale these techniques to large networks.  While these approaches attempt to scale through various means, they are not efficient enough to be used today on large real-world networks such as the wide-area networks of hyperscalers.  

This lack of scalability is fundamentally caused by a shared limitation of earlier approaches: they model and reason about network behavior \emph{monolithically}. They analyze the network configuration and routing processes as a whole, exhaustively exploring all possible control-plane behaviors induced by the complex interactions among all configuration directives and protocols. As the size of the network grows, the number of possible network states grows exponentially, limiting their ability to scale. By contrast, verification has scaled to large systems in other domains, like software or hardware, through \emph{modular} checking.  In this style, subsystems (e.g., a software function or hardware module) are verified independently to meet \emph{local} specifications (e.g., a precondition/postcondition pair) that together imply a desired global property~\cite{hoare1969axiomatic,Jones83b,conf/nato/Pnueli84}.  Prior work has used modularity to scale data-plane analysis~\cite{rcdc}, but modularizing control-plane verification is more challenging due to complex routing protocols and policies.

This paper presents a modular approach to network control plane verification. Like prior verifiers, \tool\ takes as input a network's configuration and a {\em global} property to verify.
To ensure the property, \tool\ additionally requires the user to provide local constraints that should hold on individual routers and edges. \tool\ then automatically produces a set of {\em local} checks on individual nodes and edges that \begin{enumerate*}
    \item verify the user's local constraints and
    \item ensure that these constraints imply the given end-to-end property.
\end{enumerate*}


We focus on BGP since it is ubiquitous and in many networks is the most complex process that impacts the data plane's forwarding behavior. Our approach targets two common classes of BGP reachability properties. First, \emph{safety} properties on individual routers intuitively ensure that ``bad'' routes never reach a particular node. This includes common properties like filtering bogons, preventing transit between peers, and ensuring isolation. Second, we target \emph{liveness} properties, which intuitively ensure that a ``good'' route will eventually be accepted or forwarded at a particular location. This includes many control-plane reachability queries, for example that a route received from one neighbor will be sent to another. 

Modularizing control-plane verification is challenging. Control plane behavior depends on the interaction of complex configurations with BGP, a distributed message-passing protocol.  A classical way to reason modularly about protocols is through invariants indexed by time~\cite{timepiece}, and/or employ temporal logic~\cite{lamport94}. This requires significant effort and expertise. Instead, we demonstrate that in practice a wide range of desired properties can be modularly verified without making time explicit.  Reasoning modularly about liveness properties is particularly challenging; it requires that the modular checks together imply an end-to-end path through the network.  We describe a natural approach to ensure this using two kinds of constraints:  {\em path constraints} that ensure the feasibility of a "good" path, and {\em no-interference invariants} that ensure good paths cannot be prevented.

\begin{table*}
    \centering
    \begin{tabular}{lccccc}
    \toprule
    \multirow{2}{*}{\textbf{Tool Feature}} & \multirow{2}{*}{\textbf{Minesweeper}~\cite{minesweeper}} & \multirow{2}{*}{\textbf{BagPipe}~\cite{bagpipe}} & \textbf{Plankton}~\cite{plankton} & \textbf{ARC}~\cite{ARC} & \multirow{2}{*}{\textbf{\tool}} \\ 
    & & & \textbf{Tiramisu}~\cite{tiramisu} & \textbf{Hoyan}~\cite{hoyan} \\ \midrule
    Analyzes all peer BGP routes & \CIRCLE & \CIRCLE & \Circle & \Circle & \CIRCLE  \\
    Analyzes failures & \CIRCLE & \Circle & \CIRCLE & \CIRCLE & \LEFTcircle \\
    Checks safety and liveness properties & \CIRCLE & \LEFTcircle & \CIRCLE & \CIRCLE & \CIRCLE \\
    Verification is fully automatic  & \CIRCLE & \CIRCLE & \CIRCLE & \CIRCLE & \LEFTcircle \\
    Near linear scaling with network size & \Circle & \Circle & \Circle & \Circle & \CIRCLE \\
    Localizes bugs in configurations & \Circle & \Circle & \Circle & \Circle & \CIRCLE \\
    \bottomrule
    \end{tabular}
    \caption{Comparison of prior verification tools with \tool.}
    \label{tab:related}
    \vspace{-10pt}
\end{table*}

We have formalized our approach to modular control plane verification, proved its correctness, and built a tool called \tool\ based on it. 
\tool's approach offers several advantages over the prior work, as summarized in Table~\ref{tab:related}: 

{\bf Scalability: }  \tool\ performs a {\em linear} number of checks in the network size (number of nodes and edges). Further, each check  depends only on the complexity of an individual node's configuration.  Prior approaches that reason about the \emph{joint} behavior of all nodes' policies scale at least {\em quadratically}, if not {\em exponentially}. \tool's local checks are also trivially parallelizable and enable incremental re-checking when configurations change.

{\bf Strong Guarantees: } If all of \tool's local checks are satisfied, then the specified network property is guaranteed to hold for all possible external route announcements from neighbors. Further, for safety properties our guarantees hold even in the presence of arbitrary node or link failures, though this is not true in general for liveness properties. As shown in the first two rows of the table, of the prior work only Minesweeper~\cite{minesweeper} supports reasoning about both external route announcements and failures.

{\bf Localization: }
While prior approaches identify incorrect behavior, the resulting counterexample is \emph{global}, 
making it difficult to determine which router and policy is erroneous.
By contrast, a local-check violation in \tool\ directly indicates the erroneous router and policy.

\tool's main tradeoff is that users must specify local constraints.
However, for networks designed in a modular and structured fashion, only a few simple constraints are required for any desired end-to-end property. For example, network nodes are commonly partitioned into \emph{roles}, such as border or core, each with its own responsibilities; nodes in the same role will typically have the same local constraints.

In addition, the scalability and localization properties of our tool make it easy for users to hypothesize an initial set of local constraints and then refine them iteratively based on feedback. 
We used this approach to produce the local constraints in our real-world experiments (see below), having brief discussions with the network operators based on \tool's feedback in order to either determine that an identified issue was a real configuration error or to update our local invariants appropriately.

 
We used \tool\ to verify multiple properties for BGP in a large cloud provider's wide-area network, which has hundreds of routers and tens of thousands of BGP peerings. To our knowledge no prior verification tool that reasons about all possible external route announcements has been demonstrated at this scale. 
We also ran tests on synthetic networks to show how well \tool\ scales.

In summary, we make the following contributions:

\begin{enumerate}[noitemsep]
\item 
{\bf Modularity:} A novel solution to scaling control plane verification by checking individual routers locally.

\item 
{\bf Formalization:} A formal model of BGP routing that we use to prove correctness of the modular approach.

\item 
{\bf System:} A tool \tool\ built using our approach, which has been running in a hyperscaler for six months.

\item 
{\bf Evaluation:} A demonstration of \tool's ability to scale to very large networks experimentally.

\end{enumerate}
This work does not raise any ethical issues.
\section{Approach Overview}\label{sec:example}

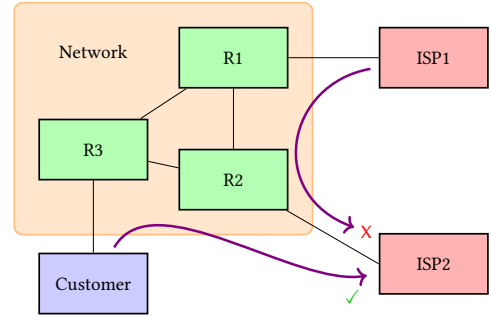
\begin{figure}[tb]
    \centering
    \scalebox{0.8}{
    \begin{tikzpicture}[
typenode/.style={},
isp/.style={rectangle, draw=black, fill=red!30, thick, minimum width=18mm, minimum height=10mm, align=center,  anchor=center},
router/.style={rectangle, draw=black, fill=green!30, thick, minimum width=18mm, minimum height=10mm, align=center,  anchor=center},
customer/.style={rectangle, draw=black, fill=blue!20, thick, minimum width=18mm, minimum height=10mm, align=center, anchor=center},
bigbox/.style={draw=orange!50, thick, fill=orange!20, rounded corners, rectangle},
]
\node[isp] (ISP1) {ISP1};
\node[isp] (ISP2) [below= 24mm of ISP1] {ISP2};

\node[router] (R1) [left= 15mm of ISP1] {R1};
\node[router]      (R2) [below= 10mm of R1]     {R2};
\node[router]      (R3)    [below left = 5mm and 5mm of R1]  {R3};
\node[customer]      (customer)   [below= 12mm of R3]  {Customer};

\node[text=red] (cross) [above left= -2mm and 0mm of ISP2] {\text{\sffamily X}};


\node[text=green!70!black] (check) [below left= -2mm and 2mm of ISP2] {\text{\sffamily \checkmark}};

\node[] (label) [above= 9mm of R3] {Network};
\draw[-] (ISP1) -- (R1);
\draw[-] (ISP2.west) -- (R2);
\draw[-] (R1) -- (R2);
\draw[-] (R1) -- (R3);
\draw[-] (R2) -- (R3);
\draw[-] (R3) -- (customer);

\draw[->, draw=violet, shorten >= 0mm, shorten <= 1.4mm, very thick, looseness=1.5] (ISP1) to [out=190, in=160] (cross);

\draw[->, draw=violet!100, shorten >= 2mm, shorten <= 1mm, very thick, looseness=0.8] (customer) to [out=60, in=190] (ISP2);


\begin{pgfonlayer}{background}
  \node[bigbox] [inner ysep=4mm,inner xsep=4mm, fit = (R1) (R2) (R3)] {};
\end{pgfonlayer}

\end{tikzpicture}







    }
    \caption{\small{Example network with safety and liveness properties (shown intuitively by the purple arrows). Routes from ISP1 should not be sent to ISP2 (safety). Routes from Customer should reach ISP2 (liveness).}}
    \label{fig:example_network}
\end{figure}

In this section we show how \tool\ works with the example in Figure~\ref{fig:example_network}. In the example network, each edge represents a connection between BGP speakers. The network contains three BGP routers: R1, R2, and R3. R1 and R2 each have an ISP as an external neighbor. R3 is connected to an external neighbor that is a customer. The network satisfies two properties. First, it satisfies the standard no-transit property that routes originating from ISP1 should not be advertised to ISP2, and second, it satisfies the property that routes from Customer, with appropriate prefixes, should eventually be sent to ISP2. The former is a \textit{safety property}, holding when a certain event never occurs, and the latter is a \textit{liveness property}, holding when a event must eventually occur. Both are network-wide policies in that they depend on the interaction of multiple routers to achieve the correct result.

Existing control-plane verifiers~\cite{minesweeper,bagpipe,ARC,tiramisu,plankton} would verify these properties by creating a representation of the possible data planes that can result from the \emph{entire network's configuration} and then searching this representation for counterexamples. This joint representation of all network node behaviors has inherent scalability limitations.

However, we observe that network configurations are highly structured and modular by design. Each router contains {\em route maps} (also called {\em route policies} and {\em route filters}), which define an import and export policy on each BGP peering session, determining which routes are rejected, which are accepted, and how accepted routes are transformed.  Each route map plays a particular role in the assurance of desired global properties.  For example, 
in the network from Figure~\ref{fig:example_network} the no-transit property can be ensured using a common approach based on communities:
\begin{enumerate*}[label=(\arabic*)]
    \item R1's import policy marks received routes from ISP1 with a BGP community (a simple 32-bit tag) with value {\tt 100:1}
    \item R2's export policy filters routes tagged with {\tt 100:1} when advertising to ISP2, and
    \item no other import or export policy strips community {\tt 100:1} from routes that it advertises.
\end{enumerate*}

Note that each of the above behaviors is \emph{node-local} and pertains to an individual BGP route map. Unlike \tool, prior control plane verification tools are not aware of this modular structure and so cannot leverage it.  Alternatively, one could envision making a tool that simply performs a set of user-specified local checks like the ones above.  However, in that case there is no guarantee that together they imply the desired end-to-end property.  Even in this simple example, the fact that it is necessary to check the third condition above is subtle and easily missed.


\begin{figure*}
    \centering
    \scalebox{0.9}{
    \includegraphics[]{figures/tikz/pipeline.tikz}
    }
    \caption{The architecture of \tool. 
    }
    \label{fig:pipeline}
\end{figure*}

Figure~\ref{fig:pipeline} shows the architecture of \tool. Like prior control-plane verifiers, it takes as input the network configuration and an end-to-end property to verify.  However, \tool\ requires the user to provide additional local constraints that capture the modular structure of the configurations. 
From these inputs \tool\ generates a set of local checks on individual nodes in the network and uses a constraint solver to verify each one.   If all of these local checks succeed, then the end-to-end property is guaranteed to hold, for all possible external route announcements from neighbors and, for safety properties, for all possible link and node failures in the network. Otherwise, \tool\ provides concrete counterexamples for each failed local check.

{\setlength{\tabcolsep}{.6em}\renewcommand{\arraystretch}{1.2}
\newcommand{\multicolwidth}{\widthof{$\Rightarrow \texttt{100:1} \in \comm{r}$}+\widthof{\textit{Routes from ISP1 are tagged with community 100:1}}+2\tabcolsep}
\begin{table*}[t!]
    \centering
    \begin{tabular}{|llll|}
    \hline
       \textbf{Type}           & \textbf{Location(s)}      & \textbf{Logical Formula} & \textbf{Description}    \\ \hline
       \rowcolor{lightblue} \text{End-to-end Property} & R2 $\rightarrow$ ISP2 & $\neg \fromisp(r)$ & \textit{No routes sent to ISP2 come from ISP1} \\ \hline
       
       \rowcolor{lightblue}                 & ISP1 $\rightarrow$ R1 & $\mathtt{True}$           & \textit{ISP1 can send our network any route} \\ \cline{2-4}
       
       \rowcolor{lightblue}                 & R2 $\rightarrow$ ISP2 & $\neg \fromisp(r)$        & \textit{No routes sent to ISP2 come from ISP1} \\ \cline{2-4}
       
       \rowcolor{lightblue}\multirow{-3}{*}{Network Invariants} & \makecell[l]{Nodes and other \\ edges in network} & \makecell[l]{$\fromisp(r)$ \\ $\Rightarrow \texttt{100:1} \in \comm{r}$} & \textit{Routes from ISP1 are tagged with community 100:1} \\ \hline
       
       \rowcolor{yellow!50}              & ISP1 $\rightarrow$ R1 & \multicolumn{2}{p{\multicolwidth}|}{$ \left( \mathtt{True} \wedge r' = \import(\mathrm{ISP1}\rightarrow \mathrm{R1}, r) \right)$
       \newline $ \Rightarrow \left( \fromisp(r') \Rightarrow \text{100:1} \in \comm{r'} \right)$}  \\ \cline{2-4}
       \rowcolor{yellow!50}              &  R2 $\rightarrow$ ISP2 & \multicolumn{2}{p{\multicolwidth}|}{$\left( \left( \fromisp(r) \Rightarrow \text{100:1} \in \comm{r} \right) \wedge r' = \export(\mathrm{R2} \rightarrow \mathrm{ISP2}, r) \right)$
       \newline $ \Rightarrow \neg \fromisp(r')$} \\ \cline{2-4}
       \rowcolor{yellow!50}              & \multirow{1}{*}{Other Edge $E$}  & \multicolumn{2}{p{\multicolwidth}|}{$\left( \left( \fromisp(r) \Rightarrow \text{100:1} \in \comm{r} \right) \wedge r' = \export(E, r) \right)$
       \newline $  \Rightarrow \left( \fromisp(r) \Rightarrow \text{100:1} \in \comm{r} \right)$} \\
       \rowcolor{yellow!50}\multirow{-5}{*}{Generated Checks} & & \multicolumn{2}{p{\multicolwidth}|}{$\left( \left( \fromisp(r) \Rightarrow \text{100:1} \in \comm{r} \right) \wedge r' = \import(E, r) \right)$
       \newline $ \Rightarrow \left( \fromisp(r) \Rightarrow \text{100:1} \in \comm{r} \right)$} \\ \hline
    \end{tabular}
    \caption{Using \tool\ to prove the no-transit property from Figure~\ref{fig:example_network}. The user-provided global property and local invariants are show in blue. \tool-generated local verification checks are shown in yellow.}
    \label{tab:example-invariants}
    \vspace{-20pt}
\end{table*}}

{\setlength{\tabcolsep}{.6em}\renewcommand{\arraystretch}{1.2}
\newcommand{\multicolwidth}{\widthof{$\Rightarrow \texttt{100:1} \in \comm{r}$}+\widthof{\textit{Routes from ISP1 are tagged with community 100:1}}+2\tabcolsep}
\begin{table*}[t!]
    \centering
    \begin{tabular}{|llll|}
    \hline
       \textbf{Type}           & \textbf{Location(s)}      & \textbf{Logical Formula} & \textbf{Description}    \\ \hline
       
       \rowcolor{lightblue} \text{End-to-end Property} &
       R2 $\rightarrow$ ISP2 & $\texttt{HasCustPrefix}(r)$ &
       \textit{Customer prefixes are advertised to ISP2} \\ \hline
       
       \rowcolor{lightblue} \text{Assumption} &
       Customer $\rightarrow$ R3 & $\texttt{HasCustPrefix}(r)$ &
       \textit{Assume customer routes are advertised to R3} \\ \hline
       
       \rowcolor{lightblue} & R3, R2, & $\texttt{HasCustPrefix}(r)$  & \textit{Routes from customer are accepted/forwarded} \\
       \rowcolor{lightblue} & R3 $\rightarrow$ R2 &  $\land \neg\text{ 100:1} \in \comm{r}$ &
       \textit{and not tagged with community 100:1} \\ 
       \cline{2-4}
       \rowcolor{lightblue}\multirow{-3}{*}{Path Constraints} & R2 $\rightarrow$ ISP2 & $\texttt{HasCustPrefix}(r)$ & \textit{Routes are forwarded to ISP2}\\ \hline
       
       \rowcolor{yellow!50}              & Customer $\rightarrow$ R3 & \multicolumn{2}{p{\multicolwidth}|}{$ \left( \texttt{HasCustPrefix}(r) \wedge r' = \import(\mathrm{Customer}\rightarrow \mathrm{R3}, r) \right)$
       \newline $ \Rightarrow \left( \texttt{HasCustPrefix}(r') \land \neg\text{100:1} \in \comm{r'} \right)$}  \\ \cline{2-4}
       
       \rowcolor{yellow!50} & & \multicolumn{2}{p{\multicolwidth}|}{$ \left(\left( \texttt{HasCustPrefix}(r) \land \neg\text{100:1} \in \comm{r} \right) \wedge r' = \export(\mathrm{R3} \rightarrow \mathrm{R2}, r) \right)$
       \newline $ \Rightarrow \left( \texttt{HasCustPrefix}(r') \land \neg\text{100:1} \in \comm{r'} \right)$} \\
       
       \rowcolor{yellow!50}              & \multirow{-2}{*}{R3 $\rightarrow$ R2}  & \multicolumn{2}{p{\multicolwidth}|}{$ \left(\left( \texttt{HasCustPrefix}(r) \land \neg\text{100:1} \in \comm{r} \right) \wedge r' = \import(\mathrm{R3} \rightarrow \mathrm{R2}, r) \right)$
       \newline $ \Rightarrow \left( \texttt{HasCustPrefix}(r') \land \neg\text{100:1} \in \comm{r'} \right)$} \\ \cline{2-4}
       
       \rowcolor{yellow!50} \multirow{-5.5}{*}{Propagation Checks} &  R2 $\rightarrow$ ISP2 & \multicolumn{2}{p{\multicolwidth}|}{$ \left(\left( \texttt{HasCustPrefix}(r) \land \neg\text{100:1} \in \comm{r} \right) \wedge r' = \export(\mathrm{R2} \rightarrow \mathrm{ISP2}, r) \right)$
       \newline $\Rightarrow \texttt{HasCustPrefix}(r')$} \\ \hline
       
       \rowcolor{yellow!50} No-interference Checks & R3, R2 & $\texttt{HasCustPrefix}(r)$ & \textit{Routes accepted at R3 and R2 with a customer} \\ 
       
       \rowcolor{yellow!50} (Safety Properties) & & $ \Rightarrow \neg\text{100:1} \in \comm{r}$ &  \textit{prefix must not have community 100:1}\\\hline

    \end{tabular}
    \caption{Using \tool\ to prove the liveness property from Figure~\ref{fig:example_network}. The user-provided global property, and path constraints are show in blue. The propagation checks are shown in yellow for the path is Customer $\rightarrow$ R3 $\rightarrow$ R2 $\rightarrow$ ISP2. The no-interference checks are safety properties proven using their own invariants (not shown).}
    \label{tab:example-liveness}
\end{table*}}
 
In the rest of this section, we show how \tool\ modularly verifies the two properties for the network in Figure~\ref{fig:example_network}. 

\subsection{Safety Properties}

{\bf End-to-end Property:}
For safety properties, the end-to-end property of interest is specified as a pair of a particular location in the network and a predicate on all routes reaching that location. Many network policies fall into this class of properties, for example bogon filtering; ensuring that a network only advertises routes to its own destinations; and forms of isolation between nodes or groups of nodes. Such properties can also express complex constraints among BGP attributes, for example that prefixes in a specific range always have a particular local preference or MED value.  

As shown in the first line of Table~\ref{tab:example-invariants}, the no-transit property specifies that no route transmitted over the edge from R2 to ISP2 should originate at ISP1.
To enable the expression of rich properties, \tool\ allows users to define \emph{ghost attributes} that conceptually update message headers with additional fields. This is a common technique in software verification, where additional variables are introduced that do not affect the computation but allow for easier property specification~\cite{ghost-var}. In the table, $\fromisp(r)$ is a boolean ghost variable that is defined by the user to be false in all originated routes, set to true by the import filter on R1 from ISP1, and left unchanged by all other filters. 

{\bf Network Invariants:} 
Users must also specify invariants that are true for routes at locations within the network. While in principle the user could specify a different invariant for each network location, many locations play the same role in the network and have the same behavior with respect to the desired end-to-end property. 
In our example, there are only three network invariants, shown in Table~\ref{tab:example-invariants}, which correspond exactly to the three node-local behaviors described earlier that ensure the no-transit property. First, no assumption is made about the routes coming from ISP1 to R1, so the associated predicate is True.  Second, routes coming from R2 to ISP2 should not come from ISP1.  Note that this invariant is identical to the end-to-end property, which is common but need not be the case.
Third, all other locations in the network should satisfy the key correctness invariant:  routes from ISP1 must be tagged with the community {\tt 100:1}.

For many safety properties, like in the example above, invariants follow a straightforward three-part structure. First, very little is assumed about routes coming from outside the network (so the associated local invariant is \texttt{True} or similarly nonrestrictive). Second, the desired global property should hold at the corresponding location in the network (the edge from R2 to ISP2 in the above example). Third, there is a key invariant that holds in the rest of the network, which intuitively describes {\em how} the network ensures the global property.  In our example above, the invariant specifies the fact that the network uses the community {\tt 100:1} to keep track of the routes that came from ISP1.  In general this invariant restricts the routes that can flow through the network to be of a limited kind, for example a specific set of prefixes or containing specific attribute values such as the MED, local preference or communities.
Notably, this three-part decomposition is analogous to the modular verification of software~\cite{hoare1969axiomatic}, which typically involves a \emph{precondition} that is assumed to hold initially, a \emph{postcondition} to be proven, and one or more \emph{inductive invariants} that hold throughout each execution and are sufficient to imply the postcondition. 

Importantly, these local invariants are far from complete specifications of the network's routing behavior.  Rather, local invariants need only describe the constraints on routes that are necessary to ensure the particular global property of interest. For example, suppose that our example network uses a route's local preference value to choose among multiple routes to a destination.  Since the local preference values don't pertain to our no-transit property, this behavior need not be specified.

{\bf Generated Checks:}
Given the invariants provided from the user, \tool\ automatically generates local checks to validate the given network invariants.  Importantly, each local check pertains to a single BGP filter on a single network router, applied to messages from a specific neighbor.  Together these checks implement a form of \emph{assume-guarantee reasoning}~\cite{Jones83b,conf/nato/Pnueli84}:  each location's network invariant is proven under the assumption that the network invariants of its directly connected locations hold.  As we prove later, together these checks imply that all local invariants in the network are respected.

Table~\ref{tab:example-invariants} shows the local checks that \tool\ automatically generates for our running example. 
The first check ensures that the import filter at R1 on the edge from ISP1 to R1 establishes the key invariant $\fromisp(r) \Rightarrow \text{100:1} \in \comm{r}$.  Since that filter tags all routes with community {\tt 100:1}, the check is easily provable by a constraint solver.  The second check ensures that the key invariant is sufficient to ensure that routes from ISP1 are not exported on the edge from R2 to ISP2.  Since the export filter at R2 on that edge drops all routes that are tagged with {\tt 100:1}, the check passes. The third set of checks ensure that the key invariant is preserved by all other import and export filters in the network.  Since these filters never strip community {\tt 100:1} from a route, the checks pass.\footnote{There are also some analogous checks for originated routes, but they are omitted here for simplicity.}
Lastly (not shown in the table), \tool\ must check that the invariant on the edge from R2 to ISP2 implies the end-to-end property.  This check is trivial since the two properties are identical.

\textbf{Output:}
If the configuration contains errors, \tool\ returns a counterexample for each local check that did not pass. In our example, suppose that R1's import filter accidentally does not add the community \text{100:1} for some routes received from ISP1.  In that case, the first generated check in Table~\ref{tab:example-invariants} would fail, producing a counterexample consisting of a concrete route that is accepted by R1 but does not get the community \text{100:1} added to it.  This counterexample directly indicates the route policy that is responsible for the error and concretely illustrates the specific local property that was violated.  Counterexamples from \tool\ are also be helpful in refining local invariants that are not precisely known in the beginning. For example, a user might write a local invariant for some network location but forget to account for a specific corner case.  In that case \tool\ will identify an ``error" due to a failed local check, and the associated counterexample informs the user how to refine that local invariant to more closely match the network location's behavior.

\subsection{Liveness Properties}

\textbf{End-to-end Property:}
For liveness properties, the end-to-end property of interest is also a pair of a particular location in the network and predicate. However, here the predicate indicates that a route satisfying the property will eventually reach that location. The property in Table~\ref{tab:example-liveness} shows that a route with a customer prefix will eventually be sent from R2 to ISP2.  If the routes of interest come from a neighbor, as in this case, then the property will only be provable under the assumption that the neighbor advertises such a route.  Users can optionally specify such an assumption, as shown in the table.

\textbf{Path and Constraints:}
As with safety properties, users need to provide a set of local constraints on individual network locations, but they take a different form for liveness properties. Users must provide a {\em path} through the network that the desired route can take to reach the destination from the source, along with local constraints for each edge and node along the path. The path does not need to be unique. Intuitively, each local constraint indicates the properties of the "good" routes that will reach that particular location, and together they constitute a witness that a "good" route will eventually reach its intended destination. As shown in Table~\ref{tab:example-liveness}, our example has two path constraints: at locations R3, R2, and R3 $\rightarrow$ $R2$ there will eventually be a route with the customer prefix that does not have the community \texttt{100:1}, and at R2 $\rightarrow$ ISP2 there will eventually be a route with the customer prefix. It is important that routes from Customer do not have the community \texttt{100:1}, or else they will be dropped at R2, due to the way that the earlier no-transit property is ensured. As described earlier, the local and concrete feedback from \tool\ can be used iteratively to identify these conditions.

\textbf{Propagation Checks:}
In order to prove the liveness property two types of checks need to be performed. First, there are local checks that together imply that a route will in fact traverse the given path, in the absence of interference from other possible paths. These checks are analogous to the generated checks for safety properties shown earlier.  Notably, in order for the first propagation check in Table~\ref{tab:example-liveness} to be satisfied, the import policy at R3 must not accept any routes tagged with community {\tt 100:1}.  One way to ensure this is for the policy to strip communities from all accepted routes.  The other two checks are straightforward.

\textbf{No-interference Checks:}
Finally, liveness properties require an additional set of checks. Since BGP only selects the best route available from all of a router's neighbors, it is not enough to show that filters do not reject "good" routes along our path. It is also necessary to show that other routes in the network can never interfere, at any node along the path. To do this, we also check that any route with the same prefix as a "good" route that can be accepted by a node on the path is also "good" --- it also satisfies the corresponding path constraint. For our example, at R3 and R2, routes with a customer prefix are checked to never have the community \texttt{100:1}. This constraint ensures that if routes for customer prefixes arrive along other paths and are preferred to those arriving on our path, those routes will still satisfy the desired property (i.e., they will be sent from R2 to ISP2).  Note that this means that our approach does not require that the specified path be unique in the network, so we can verify liveness properties even in some scenarios where there is routing redundancy.    The no-interference constraint is itself a safety property, and so in general it must be proven using the machinery shown in the previous subsection, with its own set of local invariants (not shown in the table). 

\textbf{Output:} As in the previous example, \tool\ returns a concrete counterexample for each failed propagation check and no-interference check.  For example, if R3's import policy does not properly strip the community \texttt{100:1} from accepted routes, then \tool\ will produce a concrete example illustrating this fact, allowing the user to easily understand and localize the error.

\smallskip
In summary, \tool's approach to control-plane verification leverages the modular structure that is already present in the network configurations.  By requiring the user to make this structure explicit through a set of local invariants at each location, \tool\ soundly reduces checking an end-to-end network property to a set of checks that each pertain to a single BGP import or export filter.  We formalize our approach and prove its correctness in Sections~\ref{sec:algo-safety} and \ref{sec:algo-liveness}.  

This approach has numerous benefits over the prior, monolithic approaches.  First, our approach is highly scalable, since the number of checks is linear in the number of edges in the BGP network graph.  Second, \tool's modular checks provide a very strong guarantee.  For both safety and liveness properties, the approach handles all possible external route announcements from neighbors.  For safety properties, it additionally provides resilience to arbitrary failures "for free," since it proves that "bad" routes are not received without making any assumptions about the paths that they might traverse. Third, the modular approach naturally supports incremental verification when a node is updated:  only the local checks pertaining to that node must be re-checked.  Finally, modularity has large benefits for error localization and understanding: the failure of a local check directly pinpoints the erroneous import or export filter and the local invariant that it fails to satisfy.

\section{Formal Model of BGP}\label{sec:model}

In this section we define a model of BGP in terms of traces and axioms on traces. This model is used in the next two sections to make \tool's approach precise and to prove its correctness.

\subsection{BGP Topologies and Policies}
\label{subsec:bgpModel}

We model a network's BGP configuration as consisting of two parts: a {\em topology} and a {\em policy}. A BGP network topology is a tuple of the form $(\routers , \externals, \edges)$, where:
\begin{enumerate}
    \item \routers is the set of routers for which the user provides configurations.
    \item \externals is the set of external routers. That is, there is no provided configuration, but each such router is an eBGP or iBGP peer with at least one router in \routers.
    \item \edges is the set of directed edges corresponding to BGP peering sessions.
\end{enumerate}
The network topology forms a graph with \routers $\cup$ \externals as the set of nodes and \edges as the set of edges. We will use the notation $A \rightarrow B$ to refer the directional edge ($A$, $B$) in the topology.

A BGP \emph{route} (or \emph{route advertisement}) is modeled as a tuple 
$$(\text{Prefix, ASPath, NextHop, LocalPref, MED, Comm})$$
where:
\begin{enumerate}
    \item NextHop, LocalPref, and MED are integer values
    \item Prefix is a pair consisting of an IP address and a length, both of which are integer values
    \item ASPath and Comm are lists of integer values representing the BGP path and the community tags, respectively.
\end{enumerate}
Let $\routes$ denote the set of all routes.
We will use $\comm{r}$ to refer to the Comm field of the route $r$, $\textrm{Prefix}(r)$ to refer the prefix of $r$, and so on.  Real BGP messages contain a few other attributes as well, which could be incorporated into this model. Routes can also be extended with additional ``ghost'' attributes, such as the {\tt FromISP1} attribute from Section~\ref{sec:example}. This is described in Section~\ref{subsec:ghoststate}.

We model the BGP network policy as consisting of three functions, which can be derived from the BGP and route-map configurations of each router:
\begin{enumerate}
    \item $\import : \edges \times \routes \rightarrow \routes \cup \{\reject\}$
    \item $\export : \edges \times \routes \rightarrow \routes \cup \{\reject\}$
    \item $\originate{} : \edges \rightarrow \mathbb{P}(\routes)$
\end{enumerate}
The first two correspond to the import and export route maps which are defined in the router configurations. The third models the router's ability to advertise static routes or routes from other protocols into BGP. For an edge $A \rightarrow B$ and a route $r$, $\import(A \rightarrow B, r)$ either returns the route produced when applying the import filter at $B$ to the route $r$ sent from $A$ or returns \reject if the import filter rejects the route. $\export(A \rightarrow B, r)$ either returns the route produced when applying the export filter at $A$ to the route $r$ sent to $B$ or returns \reject if the export filter rejects the route. $\originate(A \rightarrow B)$ returns the set of routes that are originated at $A$ and sent to $B$. 


\subsection{BGP Traces} \label{sec:traces}

We model the semantics of BGP as a set of allowed \emph{traces}.  Our semantics is a variant of that from the Bagpipe tool~\cite{bagpipe}.

A trace is a sequence of \emph{events}. There are three types of events that we consider: \texttt{recv}, \texttt{slct}, and  \texttt{frwd}. For $r \in \routes,\ R \text{ and } N \in \routers, \text{ and } N \rightarrow R \text{ and } R \rightarrow N \in \edges$:
\begin{enumerate}
    \item $\recv{R}{N}{r}$ occurs when $R$ receives route $r$ from neighbor $N$
    \item $\slct{R}{r}$ occurs when $R$ selects $r$ as the best route for a destination and installs it
    \item $\frwd{R}{N}{r}$ occurs when $R$ forwards route $r$ to the neighbor $N$
\end{enumerate}
We denote the set of all traces as \traces.

A \emph{valid} trace is one that could occur for a given topology and policy, according to the BGP semantics.  We formalize the notion of trace validity as a set  $\valid \subseteq \traces$ of traces that satisfy specific properties. 
We consider a trace to be valid, and hence part of the set \valid, if it satisfies a set of \textit{safety axioms}, and a set of \textit{liveness axioms}. These axioms are stated in Appendix~\ref{trace-axioms}. The safety axioms are used to prove the correctness of safety checks and state necessary conditions for an event to be in the trace. For example, if a $\texttt{slct}$ event is in the trace, then there must be a $\texttt{recv}$ event earlier and $\import$ must have transformed the received route into the selected route. The liveness axioms are used to prove the correctness of liveness checks and state sufficient conditions to show that an event occurs later in the trace. For example, if a $\texttt{slct}$ event occurs, then the result of $\export$ applied to the selected route will be used in a $\texttt{fwrd}$ event.

In our model, external neighbors can send different announcements in different traces, and events at different locations can occur in any order.

\section{Safety Verification in \tool} \label{sec:algo-safety}

In this section, we describe \tool's approach for modularly verifying safety properties and prove its correctness. 

\subsection{Inputs for Safety Checks}
\label{subsec:inputs}

\tool\ requires three inputs from the user in order to check safety properties.  The first input, the network configurations, is standard.  As described previously, the configurations are used to build the BGP topology as well as the policy functions.  

The second input is the network safety property, which requires that all route announcements that can reach a particular location satisfy certain constraints. Formally, a network safety property is a pair $(\ell, P)$ where:
\begin{equation*}
    (\ell, P) 
    \in 
    (\routers \cup \edges)
    \times \mathbb{P}(\routes)
\end{equation*}
Here $\ell$ is a location, either a router or an edge, and $P$ is a set of routes matching a particular constraint. In practice, users directly specify a logical constraint on route attributes that represents $P$. 

Each safety property $(\ell, P)$ corresponds to a property of all possible valid traces, as defined in the previous section --- all routes that can reach location $\ell$ must satisfy $P$.  Formally, a network satisfies a property $(\ell, P)$ if for all $T \in \valid, r \in \routes, R, N \in \routers, R \rightarrow N \in \edges$:
\begin{itemize}
    \item if $\ell = R$ and $\slct{R}{r} \in T$, then $r \in P$
    \item if $\ell = R \rightarrow N$ and $\frwd{R}{N}{r} \in T \lor \recv{N}{R}{r}\in T$, then $r \in P$
\end{itemize}
For example, the combination of the location $(R1 \rightarrow R2)$ and constraint $\texttt{1:1} \in \comm{r}$ together specify the property 
that if the event $\frwd{R1}{R2}{r}$ or the event $\recv{R2}{R1}{r}$ are in a valid trace, then $r$ should always have the community {\tt 1:1}. 


Finally, \tool's third input is a set of network invariants, one per location in the given network.  Formally, the network invariants are modeled as a set of pairs denoted $I$:
$$I \subseteq (\routers \cup \edges) \times \mathbb{P}(\routes)$$
Each element of the set has the form $(\ell, P)$, where $\ell$ is a location and $P$ is a set of routes, as in the network property defined above.  The semantics of each pair is a property of traces, analogous to the semantics of network properties shown above.

We require that there exist exactly one pair in $I$ per location in the given network, and we use the notation $I_{\ell}$ to denote the set $P$ of routes associated with location $\ell$ in $I$.  We also require that $I_{R\rightarrow N} = \routes$ for each edge $R \rightarrow N$ where $R \in \externals$.  In other words, we make no assumption about routes coming from external neighbors but rather assume that any route may be advertised.

\subsection{Local Checks}

Given the network configuration, network property $(\ell, P)$, and network invariants $I$, \tool\ generates the following local checks for each edge $A \rightarrow B$ in the network topology, which validate each location's network invariant using assume-guarantee reasoning:
\begin{enumerate}
    \item \textbf{Import:} For all $r, r' \in \routes$, if $r = \import(A \rightarrow B, r')$ and $r' \in I_{A\rightarrow B}$, then $r = \reject \lor r \in I_B$.
    \item \textbf{Export:} For all $r, r' \in \routes$, if $r = \export(A \rightarrow B, r')$ and $r' \in I_{A}$, then $r = \reject \lor r \in I_{A\rightarrow B}$.
    \item \textbf{Originate:} For all $r \in \routes$, if $r \in \originate(A \rightarrow B)$, then $r \in I_{A\rightarrow B}$.
\end{enumerate}
For example, the first check verifies that the import route map at $B$ on the edge $A \rightarrow B$ satisfies $I_B$, assuming that $A \rightarrow B$ satisfies its local invariant. If the router $B$ is external then the import check is not performed, and similarly if the router $A$ is external then the export and originate checks are not performed.
In our implementation of \tool, the local checks are performed by modeling import and export filters using SMT constraints and invoking an SMT solver to validate each check or provide a counterexample.

Finally, \tool\ checks that the network invariants $I$ imply the network property $(\ell, P)$.  This is done simply by requiring that $I_\ell \subseteq P$, i.e. that the network invariant for $\ell$ implies the network property $P$.  Again this check is performed with an SMT solver.

\subsection{Correctness} \label{sec:proof}

We have proven the correctness of our approach to modular safety verification.

\textbf{Theorem:} Given a BGP topology and policy, a network property $(\ell, P)$, and network invariants $I$, let $C$ be the set of Import, Export, and Originate checks that \tool\ generates.  
If all checks in $C$ pass and $I_\ell \subseteq P$, then for all $T \in \valid, r \in \routes, R, N \in \routers$:
\begin{itemize}
    \item if $\ell = R$ and $\slct{R}{r} \in T$, then $r \in P$
    \item if $\ell = R \rightarrow N$ and $\frwd{R}{N}{r} \in T \lor \recv{N}{R}{r}\in T$, then $r \in P$
\end{itemize}

\noindent
\textbf{Proof:} 
See Appendix~\ref{sec:safety-proof}.

\subsection{Ghost Attributes}\label{subsec:ghoststate}

To increase \tool's expressiveness, users can define {\em ghost attributes}, which conceptually extend each route with additional fields.  For example, the $\fromisp(r)$ ghost attribute from Section~\ref{sec:example} is used to indicate whether $r$ originated from ISP1.  A ghost attribute is defined by specifying the set of values that the attribute can take, along with updates to the \import, \export, and \originate functions that make up the given network's policy (Section~\ref{subsec:bgpModel}). 

In the case of $\fromisp(r)$ from Figure~\ref{fig:example_network}, it can be defined as a boolean attribute with the following behavior:
\begin{itemize}
    \item the import filter on ISP1 $\rightarrow$ R1 sets \fromisp to true
    \item the import filters on ISP2 $\rightarrow$ R2 and Customer $\rightarrow$ R3 set \fromisp to false
    \item other filters leave \fromisp unchanged
    \item all originated routes have \fromisp set to false
\end{itemize}

Other natural network properties can be expressed using ghost attributes. For example, a \waypoint attribute that is true only for routes processed by a particular router $R$  can be defined by specifying that filters on $R$ set \waypoint to true, origination as well as import filters from external neighbors at other routers set \waypoint to false, and all other filters in the network leave \waypoint unchanged. 

Ghost attributes do not affect the description of \tool\ or proof of its correctness above, as they do not depend on the specific set of attributes that are in a route.

\subsection{Fault Tolerance for Safety Properties}
\label{subsec:failures}

A significant benefit of \tool's approach to control-plane verification of safety properties is that it supports reasoning about failures ``for free.'' That is, if all of \tool's checks pass, then the given network property is guaranteed to hold not only in the failure-free case but also in the presence of \emph{arbitrary} node and link failures.  

\tool\ soundly reasons about failures because of our over-approximate notion of trace validity (Section~\ref{sec:traces} and Appendix~\ref{trace-axioms}). Specifically, any trace that is feasible according to the given BGP topology and passes the import and export filters along the corresponding path is considered valid. Hence, every trace that can occur under any failure scenario is already considered valid.  By our correctness theorem, all of these traces satisfy the property $(\ell, P)$. 
\section{Liveness Verification in \tool}\label{sec:algo-liveness}

We now describe how \tool\ checks liveness properties modularly. Proving liveness properties modularly is more difficult than proving safety properties, since it requires showing both that "good" routes are allowed and that interfering routes are not.  

\subsection{Inputs for Liveness Checks}

The inputs for a liveness check consist of the following: 
\begin{enumerate}
    \item The network configurations
    \item A liveness property $(\ell, P) \in \{\routers \cup \edges\} \times \mathbb{P}(\routes)$
    \item A path $(\ell_1, \dots, \ell_n = \ell)$ where $\ell_i \in \{\routers \cup \edges\}$
    \item A constraint $C_1 \dots C_n$ for each location in the path, where $C_i \in \mathbb{P}(\routes)$
    
    
\end{enumerate}
The property $(\ell, P)$ represents a liveness property of all valid traces, namely that there will eventually be a route at $\ell$ that satisfies $P$. Formally, this means for all $T \in \valid$, either:
\begin{itemize}
    \item $\ell \in \routers$ and there exists $r'$ such that $\slct{\ell}{r'} \in T$ and $P(r')$ holds, or
    \item $\ell \in \edges$ and there exists $r'$ such that $\texttt{frwd}(\ell, r') \in T$ and $P(r')$ holds
\end{itemize} 

The path $(\ell_1, \ell_2, \dots, \ell_{n-1},  \ell_n = \ell)$ is a sequence of routers and edges that we expect the route to travel across. We require that it represents an actual topological path in the network: if $\ell_i = R \in \routers$ then for some $N$, $\ell_{i+1} = R \rightarrow N$, and if $\ell_i = R \rightarrow N$, then $\ell_{i+1} = N$. For example, ISP1 $\rightarrow$ R1, R1, R1 $\rightarrow$ R3, R3, R3 $\rightarrow$ Customer is a path in the network from Figure~\ref{fig:example_network}. The last location $\ell_n$ must be the location $\ell$ of the end-to-end property that we are verifying.


The constraints $C_1 \dots C_n$ are properties that represent the set of "good" routes that reach each $\ell_i$ along the path. They play a role analogous to the local invariants $I_{\ell_i}$ for proving safety properties, described earlier.  The property $C_1$ for the first location in the path is simply assumed to hold; in practice it is usually an edge coming from an external router, in which case it is not possible to prove.  Rather, the best we can do is prove that if that router sends a "good" route, then it will eventually reach its intended destination in the network. 

\subsection{Local Checks}\label{no-interference}

The checks for liveness can be broken up into two parts: checks that prove propagation along the given path, and checks that prove there is no interference from outside routes. 

\textbf{Propagation along a path:}
These checks are analogous to the local checks performed for safety verification, but they are only checked along the given path. Together they ensure that the import and export filters along the path $(\ell_1, \dots, \ell_n)$ do not drop "good" routes.  Specifically, for all valid traces $T$ and $i < n$:

\noindent
If $\ell_i = R \in \routers$, then: 
\begin{align*}
    &C_i(r) \land r' = \export(R \rightarrow N, r) \\ &\implies r' \ne \reject \land C_{i+1}(r')
\end{align*}
and if $\ell_i = R \rightarrow N \in \edges$, then: 
\begin{align*}
    &C_i(r) \land r' = \import(N \rightarrow R, r) \\ &\implies r' \ne \reject \land C_{i+1}(r')
\end{align*}


\textbf{No interference:}
Next, we need to verify that it is not possible for a router along the path to select a "bad" route with the same prefix as a "good" route. 
Let $\texttt{Prefix}(C_i)$ refer the set of prefixes with at least one route in $C_i$:
\begin{equation*}
    \{p\ |\ p = \texttt{Prefix}(r) \land r \in C_i\}
\end{equation*}
Then at each router $\ell_i$ along the path we must prove the following safety property:
\begin{equation*}
    (\ell_i, \texttt{Prefix}(r) \in \texttt{Prefix}(C_i) \implies C_i(r))
\end{equation*}
These properties can be proven using our existing approach for proving safety properties (Section~\ref{sec:algo-safety}), given appropriate local invariants.

\textbf{Implying the network property:} The above checks ensure that all of the local $C_i$ constraints in fact hold. Finally, \tool\ generates a local check that $C_n \subseteq P$, similar to the analogous check for safety properties, to ensure that the local constraints imply the desired end-to-end liveness property.

\subsection{Correctness}

We have proven the correctness of our approach to modular safety verification.

\textbf{Theorem:} Given the following:
\begin{itemize}
    \item The network configurations
    \item A liveness property $(\ell, P)$
    \item A path $S = (\ell_1, \ell_2, \dots, \ell_{n-1}, \ell_n = \ell)$
    \item A constraint for each location $C_1 \dots C_n$
\end{itemize}
For all valid traces $T$, if all of the following are true:
\begin{enumerate}
    \item all checks (propagation, no interference) pass
    \item there exists $r$ such that $\texttt{recv}(\ell_1, r) \in T \land C_1(r)$
    \item $C_n \subseteq P$
    \item there are no link failures along the path 
\end{enumerate}
then there exists $r'$ such that either:
\begin{itemize}
    \item $\ell \in \routers$ and there exists $r'$ such that $\slct{\ell}{r'} \in T$ and $P(r')$ holds, or
    \item $\ell \in \edges$ and there exists $r'$ such that $\texttt{frwd}(\ell, r') \in T$ and $P(r')$ holds
\end{itemize} 
\noindent
\textbf{Proof:}
See Appendix~\ref{sec:liveness-proof}.

\medskip
\noindent
Notably, the correctness only depends on there being no link failures along the given path, so the property holds even if there are failures elsewhere.

\section{Evaluation}
\label{sec:evaluation}

\subsection{Cloud WAN}

\begin{table*}[ht!]
    \normalsize
    \begin{subtable}[]{\textwidth}
        \centering
        \setlength{\tabcolsep}{9.8pt}
        \begin{tabular}{|l|lll|}
        \hline
           \textbf{Type}           & \textbf{Locations ($l$)}      & \textbf{Logical Formula ($I_l$)} & \textbf{Description}    \\ \hline
           \makecell[l]{End-to-end \\ Property} & 
           \makecell[l]{Any R in \\ network} & 
           \makecell[l]{$FromPeer(r) \implies$ \\ $\textsc{Prefix}(r) \notin \textsc{Bogons }$}& 
           \makecell[l]{\textit{Bogon prefixes from peers should not be accepted}} \\ \hline
           
            & $R\in \routers$ & \makecell[l]{$FromPeer(r) \implies$ \\ $ \textsc{Prefix}(r) \notin \textsc{Bogons }$ }& \textit{Bogon prefixes from peers should not be accepted at routers} \\ \cline{2-4}
            \makecell[l]{Network \\ Invariants}
            & \makecell[l]{Internal edges \\ R1 $\rightarrow$ R2}& \makecell[l]{$FromPeer(r) \implies$ \\ $ \textsc{Prefix}(r) \notin \textsc{Bogons }$ }& \textit{Bogon prefixes from peers should not be sent along edges} \\ \cline{2-4}
            & Other & $True$ & \textit{Edges to and from external peers are unconstrained} \\ \hline
        \end{tabular}
        \caption{End-to-end property and network invariants needed to verify that the network does not accept bogons from external peers.}
        \label{tab:bogon-invariants}
    \end{subtable}
    
    \begin{subtable}[]{\textwidth}
        \centering
        \setlength{\tabcolsep}{8.2pt}
        \begin{tabular}{|l|llp{2.6in}|}
        \hline
           \textbf{Type}           & \textbf{Locations ($l$)}      & \textbf{Logical Formula ($I_l$)} & \textbf{Description}    \\ \hline
           \makecell[l]{End-to-end \\ Property} &  $R \notin \textsc{Region}$ & \makecell[l]{$FromRegion(r) \implies$ \\ $\textsc{ Prefix}(r) \notin \textsc{ReusedIPs}$}& 
           \makecell[l]{\textit{Routers outside a region should not accept} \\ 
           \textit{routes with reused addresses from that region}} \\ \hline
           
            & $R \in \textsc{Region}$ & \makecell[l]{$FromRegion(r)\ \land$ \\ $\textsc{ Prefix}(r) \in \textsc{ReusedIPs } \implies $ \\ $ \textsc{RegionalComms} \cap \comm{r} = \{C\}$ }& \textit{Routes with reused addresses are tagged with a community for that region and no other region} \\ \cline{2-4}
            
            \makecell[l]{Network \\ Invariants} & $R \notin \textsc{Region}$ & \makecell[l]{$FromRegion(r) \implies $ \\ $\textsc{ Prefix}(r) \notin \textsc{ReusedIPs}$} & \makecell[l]{\textit{Routers outside a region should not accept} \\\textit{routes with reused addresses from that region}} \\ \cline{2-4}
            
            & R1 $\rightarrow$ R2 & $I_{R1}$ & \textit{Edges have same invariant as sending router} \\ \cline{2-4}
            & E $\rightarrow$ R & $\comm{r} = \emptyset$ & \textit{Routes from external peers have no communities} \\ \hline
        \end{tabular}
        \caption{End-to-end property and network invariants needed to verify that reused addresses are not accepted by any router outside the region. }
        \label{tab:reuse-invariants}
    \end{subtable}

    \begin{subtable}[]{\textwidth}
        \centering
        \setlength{\tabcolsep}{6.6pt}
        \begin{tabular}{|l|llp{2.6in}|}
        \hline
           \textbf{Type}           & \textbf{Locations ($l$)}      & \textbf{Logical Formula ($I_l$)} & \textbf{Description}    \\ \hline

           \makecell[l]{End-to-end \\ Property} &  $R_2 \in \textsc{Region}$ & \makecell[l]{$FromRegion(r)\ \land$ \\ $\textsc{ Prefix}(r) \in \textsc{ReusedIPs}$} & \makecell[l]{
           \textit{$R_2$ inside a region eventually accepts a route} \\ \textit{with reused addresses from that region}}
           \\ \hline
           
           \makecell[l]{Assumption} & \makecell[l]{Edge from data \\ center $D \rightarrow R_1$} & \makecell[l]{$FromRegion(r)\ \land$ \\ $\textsc{ Prefix}(r) \in \textsc{ReusedIPs}$} & \makecell[l]{\textit{Assume there is a route from the data center to} \\ \textit{$R_1$ with a reused prefix}}
           \\ \hline
           
            \makecell[l]{Path \\ Constraints} & $R_1, R_2, R_1 \rightarrow R_2$ & \makecell[l]{$FromRegion(r)\ \land $ \\ $\textsc{ Prefix}(r) \in \textsc{ReusedIPs}\ \land$ \\ $\textsc{RegionalComms} \cap \comm{r} = \{C\}$} & \makecell[l]{\textit{$R_1$ and $R_2$ eventually select a route with reused} \\ \textit{prefixes and the regional community}}
            \\ \hline
        \end{tabular}
        \caption{End-to-end property and path constraints needed to verify that reused addresses are eventually selected by each WAN router in that region. We assume that the route flows from the data center along the path $D \rightarrow R_1 \rightarrow R_2$.}
        \label{tab:reuse-liveness}
    \end{subtable}

    \caption{End-to-end properties and network invariants for three use cases in the WAN.}
    \label{tab:eval-invariants}
    \vspace{-10pt}
\end{table*}

We used \tool\ to modularly verify properties of the wide-area network (WAN) of a major cloud provider, containing hundreds of routers and tens of thousands of peering sessions. In doing so, we show that:
\begin{enumerate*}[label=(\arabic*)]
    \item important behavioral properties in real-world networks can be expressed in \tool;
    \item these properties can be proven through a combination of modular checks;
    \item this approach scales, allowing properties to be verified quickly; and
    \item if a local check does not succeed, it produces actionable information,  indicating a bug in either a specific route map or a specific local invariant.
\end{enumerate*}
To our knowledge no prior tool that verifies properties of all possible external announcements from neighbors has been demonstrated to scale to such a size. 

We used \tool\ to verify two classes of properties that the wide-area network must satisfy.  In all cases we determined the intended network behavior by inspecting the configurations and talking with the network operators, and the local constraints were written based on that intent. This process was typically iterative. That is, we would write an initial property specification and its set of local invariants based on our current understanding of how the network operates. If \tool\ reported violations of local checks, we would inspect the counterexamples and discuss with operators, either determining that the bugs are real errors or identifying special cases that led to refined local invariants and (sometimes) refined end-to-end property specifications.

\textbf{Implementation:} We implemented \tool\ as a tool in C\#. The tool parses and extracts the BGP policy along with import and export route maps from each configuration, while supporting common attributes of BGP routes such as communities, AS path, MED, local preference, along with common route map features, like matching on and setting attributes. The tool allows users to provide local invariants written as a C\# function using the Zen constraint solving library~\cite{zenlib}, and to specify the routers and policies of interest. The Zen library translates the functions into SMT formulas that are solved by Z3~\cite{z3}. For each local check that fails, the tool returns a counterexample consisting of a specific route map and a concrete input route that leads to a violation.





{\bf Internet Peering Policies:}
We used \tool\ to verify that 11 different kinds of "bad" routes are never accepted from peers.  Each of these properties can be expressed as a safety property on each node $R$ in the network of the following form:
$$ (R, \{r\ | FromPeer(r) \implies Q(r)\})$$
with different properties $Q(r)$. These include properties like not accepting bogons or routes with invalid AS paths. An example of the invariants for the no-bogons property is shown in Table~\ref{tab:bogon-invariants}. 
The network has a set of Internet {\em edge routers}, that peer with Internet service providers, other cloud providers, and customers, and so act as gateways between the cloud provider and the Internet. The wide-area network ensures that "bad" routes are not admitted by filtering them at all of the Internet edge routers.

As mentioned earlier, running \tool\ to check these properties is an iterative process, which involves refining the local constraints based on operator feedback.  In the end, through this process
\tool\ identified 11 actual configuration errors.  
These included cases where a route map denied more traffic than intended, and inconsistencies between the filters of edge routers that are intended to have similar behavior. 
For example, in one case, among the hundreds of similarly defined peering sessions, it was discovered that a handful had ad-hoc policies that filtered AS paths differently.
All of the findings were latent bugs that did not have an immediate impact, but could become impactful in the presence of failures or changes in the external announcements received from neighbors. Further, because \tool\ is sound the operators can be sure that these are the {\em only} violations of the desired end-to-end properties. As of this writing, all identified errors are prioritized for fixing by network engineers.

Verification with \tool\ is highly scalable.  The maximum time that it took \tool\ to sequentially run all of the local checks for any single property was 15 minutes, across all devices in the network.  As another data point, an automation that sequentially ran the local checks for four of the properties across all of the hundreds of edge routers took a total of 16 minutes. Given that each of these checks can be run independently on each device configuration, it would also be easy to parallelize these checks in the future in order to scale horizontally for a large number of devices.

While using \tool\ to verify these 11 properties, we also learned best practices for writing properties. Initially, we combined multiple properties into a single property for \tool\ to check.  However, we found that writing multiple simpler properties, with associated simpler local constraints, was not only easier to write and debug but also was usually faster to run, since the constraints are simpler for the underlying SMT solver to process.

{\bf Proper IP Reuse:} In the second use case, \tool\ verified proper usage of reused IPs within the network. The cloud network is partitioned into dozens of \emph{regions}, and some private IPv4 addresses are reused in different regions. There is a safety property that traffic sent to these private addresses must stay within the region, and also a liveness property that routes to reused addresses are advertised to other WAN routers in the same region. We verified both of these properties for all regions in the network.

The safety property to verify is as follows, for each router $R$ that is \emph{not} part of the region of interest:
$$(R, \{r\ |\ FromRegion(r) \implies \textsc{ Prefix}(r) \notin \textsc{ReusedIPs}\})$$
Here $FromRegion(r)$ is a ghost variable that is set to true only on routes coming from external routers in the particular region, and \textsc{ReusedIPs} is the set of prefixes that are reused.
The liveness property requires that in each region, a route with a reused prefix from the data center routers can reach all other routers in that region, possibly going through one intermediate router. That is, for every pair of WAN routers $R_1$ and $R_2$ in the same region, if $R_1$ is connected to a data center router $D$, then routes with a reused prefix can travel  $D \rightarrow R_1 \rightarrow R_2$.

The WAN enforces these properties by tagging routes for reused IP addresses with a region-specific community $C$ when they are received from data centers.  Routers in the same region then accept routes tagged with that community, while routers in other regions reject them.
The local constraints we used to verify the safety and liveness properties are shown in Table~\ref{tab:reuse-invariants} and \ref{tab:reuse-liveness} respectively.  One subtlety is that routes to reused IP addresses in the region of interest must not only have the community $C$, but they also must not be tagged with any other region's community.  Otherwise, these routes could be accidentally accepted by other regions.  The local constraints validate this property, and the WAN enforces it by deleting all communities on routes coming from the data centers, before adding the community $C$. 

The communities used in each region were documented in a  metadata file, which made it easy for us to write the local constraints for each region. In one case, \tool\ found a violation where a router used a community that was not present in the metadata file. The operators acknowledged that this was a bug that could cause some traffic to be redirected. \tool\ was able to verify all other local checks, for both the safety and the liveness properties.

\subsection{Scaling Experiments}


To illustrate the scaling benefits of modular checking, we compared \tool\ with Minesweeper~\cite{minesweeper} on synthetic test cases. For a fair comparison, we created an implementation of \tool\ that is built on top of the same parser and constraint generation system as Minesweeper. This is a different implementation from the one used on the cloud network. 
We use a BGP full mesh where each router is connected to one external neighbor through eBGP and all other routers through iBGP. This leads to a total of $N^2$ edges in a network of size $N$. The network's configuration is relatively simple, with each eBGP connection using only prefix and community filters. We checked a no-transit safety property, similar to the example in Figure~\ref{fig:example_network}.

\pgfplotstableread[row sep=\\,col sep=&]{
    Routers & 
    Edges & 
    Time &
    Variables &
    Assertions &
    CheckTime &
    MinesweeperTime &
    MinesweeperVariables &
    MinesweeperAssertions &
    MinesweeperCheckTime 
    \\ 
    10  & 100 & 
    4 & 1809 & 778 & 0.6 &
    4.282 & 27327 & 43319 & 2.899
    \\
    20  & 400 & 
    13 & 3329 & 1408 & 3 &
    340.273 & 153952 & 240154 & 334.507
    \\
    30  & 900 & 
    26 & 3361 & 1401 & 6 &
    580.942 & 432977 & 679589 & 565.770
    \\
    40  & 1600 & 
    50 & 3529 & 1448 & 12 & 0
     & 926702 & 1459924  & 0
    \\
    50  & 2500 & 
    74 & 3561 & 1441 & 18 & 0
     & 1695127 & 2677159 & 0
    \\
    60  & 3600 & 
    114 & 3729 & 1488 & 29 & 0
     & 2798252 & 4427294 & 0
    \\
    70  & 4900 & 
    152 & 3761 & 1481 & 39 & 0
      & 4296077 & 6806329 & 0
    \\
    80  & 6400 & 
    205 & 3929 & 1528 & 52 & 0
      & 6248602 & 9910264 & 0
    \\
    90  & 8100 & 
    256 & 3961 & 1521 & 62 & 0
     & 8715827 & 13835099 & 0
    \\
    100 & 10000 & 
    327 & 4129 & 1568 & 82 & 0
      & 11757752 & 18676834 & 0
    \\
}\mydata

\pgfplotsset{compat=1.3}
\begin{figure*}[tb!]
    \captionsetup{margin=2pt,}
    \subcaptionbox{Number of variables and constraints generated by Minesweeper for synthetic networks.\label{fig:mine_sizes}}
    {
    \scalebox{0.8}{
        \begin{tikzpicture}
            \begin{axis}[
                    ybar,
                    xtick=data,
                    bar width=.15cm,
                    width=\columnwidth,
                    height=4cm,
                    legend style={at={(0.5,1)},
                        anchor=south,legend columns=1},
                    ylabel={Count},
                    xlabel={Number of Routers},
                    ymin=0,
                    ymax=22000000,
                    yticklabel style={
                       /pgf/number format/fixed,
                    },  
                    scaled y ticks=false
                ]
                \addplot table[x=Routers,y=MinesweeperVariables]{\mydata};
                \addplot table[x=Routers,y=MinesweeperAssertions]{\mydata};
            \legend{Number of Variables (left), Number of Constraints (right)}
            \end{axis}
        \end{tikzpicture}
        }
    }%
    \hfill
    \subcaptionbox{The maximum number of variables and constraints in any single local check generated by \tool.\label{fig:light_sizes}}
    {
    \scalebox{0.8}{
        \begin{tikzpicture}
            \begin{axis}[
                    ybar,
                    xtick=data,
                    bar width=.15cm,
                    width=\columnwidth,
                    height=4cm,
                    legend style={at={(0.5,1)},
                        anchor=south,legend columns=1},
                    ylabel={Count},
                    xlabel={Number of Routers},
                    ymin=0,
                    yticklabel style={
                       /pgf/number format/fixed,
                    },  
                    scaled y ticks=false
                ]
                \addplot table[x=Routers,y=Variables]{\mydata};
                \addplot table[x=Routers,y=Assertions]{\mydata};
            \legend{Maximum Number of Variables (left), Maximum Number of Constraints (right)}
            \end{axis}
        \end{tikzpicture}
        }
    }
    \subcaptionbox{Time used by Minesweeper to verify a property of synthetic networks. Runtime for networks with 40 routers or more exceeds two hours, not including time to parse configurations.
    \label{fig:mine_times}}
    {
    \scalebox{0.8}{
        \begin{tikzpicture}
            \begin{axis}[
                    ybar,
                    xtick={10, 20, 30, ..., 100},
                    bar width=.15cm,
                    width=\columnwidth,
                    height=4cm,
                    legend style={at={(0.5,1)},
                        anchor=south,legend columns=1},
                    ylabel={Time (seconds)},
                    xlabel={Number of Routers},
                    ymin=0,
                    ymax=650,
                    yticklabel style={
                       /pgf/number format/fixed,
                    },  
                    scaled y ticks=false
                ]
                \addplot table[x=Routers,y=MinesweeperCheckTime]{\mydata};
                \addplot table[x=Routers,y=MinesweeperTime]{\mydata};
                
                \node[] at (axis cs: 70, 100) {Exceeds 2hrs};
            \legend{Constraint Solving Time (left), Total Time (right)}
            \end{axis}
        \end{tikzpicture}
        }
    }%
    \hfill
    \subcaptionbox{Time used by \tool\ to verify a property in synthetic networks, not including time to parse configurations.
        \label{fig:light_times}}
    {
    \scalebox{0.8}{
        \begin{tikzpicture}
            \begin{axis}[
                    ybar,
                    xtick={10, 20, ..., 100},
                    ytick={200, 400, 600},
                    bar width=.15cm,
                    width=\columnwidth,
                    height=4cm,
                    legend style={at={(0.5,1)},
                        anchor=south,legend columns=1},
                    ylabel={Time (seconds)},
                    xlabel={Number of Routers},
                    ymin=0,
                    ymax=650,
                    yticklabel style={
                       /pgf/number format/fixed,
                    },  
                    scaled y ticks=false
                ]
                \addplot table[x=Routers,y=CheckTime]{\mydata};
                \addplot table[x=Routers,y=Time]{\mydata};
            \legend{Constraint Solving Time (left), Total Time (right)}
            \end{axis}
        \end{tikzpicture}
        }
    }
    
    \caption{Comparing \tool\ and Minesweeper on synthetic networks of various sizes.}
    \label{fig:graphs}
\end{figure*}
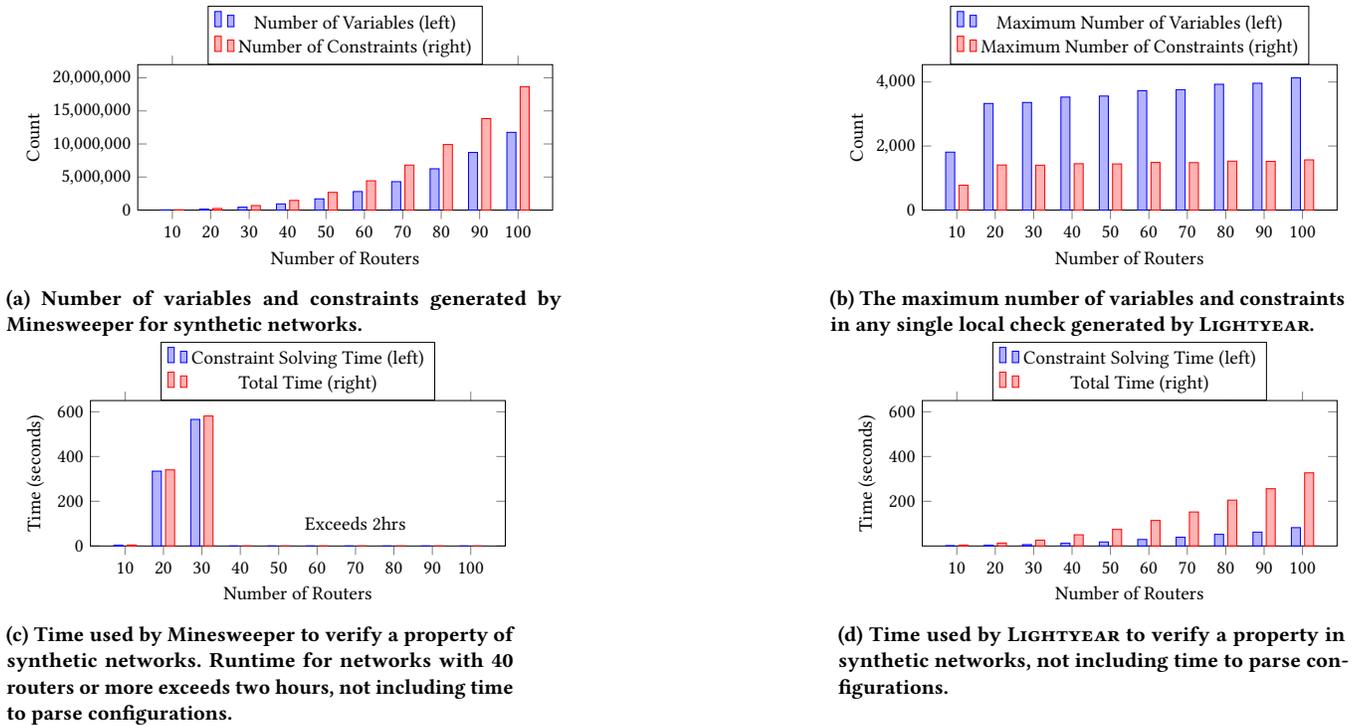

Figure~\ref{fig:graphs} provides details on these results by comparing the number of SMT variables and constraints generated by each tool, as well as the amount of time used to solve the SMT constraints compared to the total computation time.  
As the network size increases, Minesweeper requires several orders of magnitude more SMT variables and constraints than the maximum number required by \tool\ for any local check (compare Figures~\ref{fig:mine_sizes} and \ref{fig:light_sizes}).  As a result, SMT solving time dominates the run time of Minesweeper and is the limiting factor on its ability to scale, while for \tool\ the solving time is a relatively small portion of the total time (compare Figures~\ref{fig:mine_times} and \ref{fig:light_times}). Minesweeper does not terminate within two hours when run on a network of size 40, while \tool\ verifies a network of size 100 in 5.5 minutes.



\section{Related Work}
\label{sec:relatedwork}

{\bf Control Plane Verification:} State-of-the-art approaches to network control-plane verification were summarized in Table~\ref{tab:related}.  Unlike \tool, these approaches are all {\em monolithic} --- they require joint analysis of the configurations of all nodes --- which dramatically limits scalability. Compared to \tool, Minesweeper's worst case complexity is exponential in the network size. Other improvements not only
reduce generality but are at least quadratic in the network size even when using specialized algorithms. Most approaches make tradeoffs in expressiveness, for example giving up the ability to reason about all possible BGP announcements from neighbors~\cite{ARC,plankton,tiramisu,ye20accuracy}.  In contrast, \tool's modular approach only requires reasoning about individual BGP route maps in isolation and so is highly scalable.  \tool\ also provides guarantees across all possible external announcements and, for safety properties, arbitrary failures. 

{\tt rcc}~\cite{rcc} validates important properties of BGP configurations, largely through local checks on individual configuration. However, {\tt rcc} is limited to specific "best practice" policies, and
there is no guarantee that the local checks together ensure the desired end-to-end properties.

Closest to our work are recent techniques for modular control-plane verification, Kirigami~\cite{kirigami} and Timepiece~\cite{timepiece}, which also use assume-guarantee reasoning for the control plane via local invariants.  However, each approach makes a different set of tradeoffs than \tool.  Kirigami's local invariants require the exact routes that will arrive on a particular edge. Because these invariants are fully concrete, Kirigami cannot reason about arbitrary route announcements from neighbors or give guarantees in the presence of failures.

Timepiece allows for expressive local invariants and properties, using an explicit notion of time.  In Timepiece, routing protocols have discrete, synchronized time steps, and in each step, each router computes the best route among those it receives. This model allows Timepiece to specify and check temporal-logic properties but requires users to provide complex local invariants for each node that are explicitly indexed by time.  In our model routes can be sent and arrive in arbitrary orders, and we demonstrate how to specify and check common safety and liveness properties without explicit time. 

Another line of work has improved scalability of control-plane verification through forms of {\em abstraction}~\cite{bonsai,shapeshifter}: the full network is analyzed monolithically, but irrelevant or redundant configuration information is abstracted away to simplify the analysis.  Our work is orthogonal to this line of work; the two approaches could be combined. 

{\bf Data Plane Verification:} Other tools check properties of forwarding state, rather than network configurations~\cite{anteater,HSA,Veriflow,yang2015real,secguru,netkat,nodpaper}.  These approaches generally require joint reasoning about the entire network. A recent exception is RCDC~\cite{rcdc}, which modularly verifies global reachability contracts in a data center via local checks.  However, RCDC is specific to one data center design and does not provide a general framework for decomposing global property checks into local checks. Another approach~\cite{PlotkinBLRV16} exploits abstraction, such as symmetries, to scale data-plane verification.

{\bf Modular Verification:} 
Assume-guarantee reasoning~\cite{Jones83b, conf/nato/Pnueli84} enables modular verification in other domains.
A global property is modularized by providing each system component with local invariants that it must satisfy, assuming other components satisfy their invariants. \tool\ applies this methodology to networks to generate the local checks that each BGP policy must satisfy.

Verification often requires identifying \emph{inductive invariants}, properties that hold over some unbounded space of system states, such as the iterations of a loop~\cite{hoare1969axiomatic}.
Such invariants arise naturally in networks and enable many locations to use the same local invariant. Typically, a small set of nodes establishes an inductive invariant (e.g., by attaching a community), and this invariant holds through the network as long as other nodes ``do no harm'' (e.g., never remove communities).

\section{Conclusion}

Exploiting symmetries in network verification~\cite{PlotkinBLRV16} is natural because of hardware design patterns such as fat trees. Similarly, exploiting modularity in control plane verification is natural because of design patterns in the way configurations are written and maintained in well engineered networks.  We have confirmed this hypothesis in six months of deployment at a major cloud vendor.  Further, \tool\ finesses the need to reason about time to prove safety and liveness, offering a sweet spot between expressiveness 
and complexity that has worked well for many desired properties in our network.


In \tool, users must provide local network constraints.  While in our experience it has been easy to determine these constraints, we believe it is possible to instead \emph{learn} local invariants automatically from configurations in the future, for example when properties are enforced via communities.  



\bibliographystyle{ACM-Reference-Format}
\bibliography{reference}

\appendix

\medskip

\noindent
Appendices are supporting material that has not been peer-reviewed.

\section{BGP Trace Axioms}\label{trace-axioms}

The safety axioms consist of the following properties, for all $1 \leq k \leq n$:
\begin{enumerate}
    \item If $A_k = \recv{R}{N}{r}$, then either:
    \begin{enumerate}
        \item $N \in \externals$, or
        \item there exists $j < k$ such that $A_j = \frwd{N}{R}{r}$
    \end{enumerate}
    \item If $A_k = \slct{R}{r}$, then there exists $j < k$, $r' \in \routes$, and $N \in \routers \cup \externals$ such that $A_j = \recv{R}{N}{r'}$ and $r = \import(N \rightarrow R, r')$
    \item If $A_k = \frwd{R}{N}{r}$, then either:
    \begin{enumerate}
        \item $r \in \originate(R \rightarrow N)$, or
        \item there exists $j < k$ and $r' \in \routes$ such that $A_j = \slct{R}{r'}$ and $r = \export(R \rightarrow N , r')$
    \end{enumerate}
\end{enumerate}

The liveness axioms depend on the BGP route preference relation, which selects routes to the same prefix by comparing their local preference, AS paths, and other attributes. We say that $r_1 > r_2$ if $r_1$ is preferred over $r_2$.  The liveness axioms
consist of the following properties, for all $1 \leq k \leq n$:

\begin{enumerate}
    \item If all of the following are true:
    \begin{itemize}
        \item $A_k = \slct{R}{r} \in T$
        \item $r' = \export(R \rightarrow N, r)$ with $r' \ne \reject$
    \end{itemize}
    then there exists $j > k$ such that $A_j = \frwd{R}{N}{r'}$
    \item If $r \in \originate(R \rightarrow N)$ then there exists $j > k$ such that $A_j = \frwd{R}{N}{r}$
    \item If $A_k = \frwd{R}{N}{r}$ and there is no link failure along $R\rightarrow N$, then there exists $j > k$ such that $A_j = \recv{R}{N}{r}$
    \item If all of the following are true:
    \begin{itemize}
        \item $A_k = \recv{R}{N}{r}$
        \item $r' = \import(N \rightarrow R, r)$ with $r' \ne \reject$
        \item For all neighbors $N' \neq N$ and routes $r''$: \\ if $\texttt{Prefix}(r) = \texttt{Prefix}(r'')$ and $\recv{R}{N'}{r''} \in T$, then $r' > \import(N \rightarrow R, r'')$
    \end{itemize}
    then there exists $j > k$ such that $A_j = \slct{R}{r'} \in T$. 
\end{enumerate}

\section{Correctness Proof for Safety}\label{sec:safety-proof}

In this section we prove that \tool's modular approach to control-plane verification is correct.

First we state and prove the key lemma, which says that the local checks are sufficient to ensure that the network invariants $I$ hold, for all valid traces.

\medskip
\noindent
\textbf{Lemma:} Given a BGP topology and policy as well as network invariants $I$, let $C$ be the set of Import, Export, and Originate checks that \tool\ generates.  If all checks in $C$ pass, then for all $T \in \valid, r \in \routes, R, N \in \routers$:
\begin{itemize}
    \item if $\slct{R}{r} \in T$, then $r \in I_R$
    \item if $\frwd{R}{N}{r} \in T \lor \recv{N}{R}{r}\in T$, then $r \in I_{R\rightarrow N}$
\end{itemize}

\noindent
\textbf{Proof:} The proof is by induction on the length of the (partial) trace $T$.

\noindent
\textbf{Base case:} For a partial trace of length 0, there are no events, so the statement is vacuously true.

\noindent
\textbf{Inductive case:} Suppose $T = A_1, A_2, \dots, A_{k+1}$. We assume by induction that the statement is true for $A_1, A_2, \dots, A_k$. We do a case analysis on the event $A_{k+1}$:

\noindent
Case $A_{k+1} = \recv{R}{N}{r}$, so we have to show that $r \in I_{N\rightarrow R}$. By the trace validity axioms, either:
\begin{enumerate}
    \item $N \in \externals$. In this case we know that $I_{N\rightarrow R} = \routes$, so $r \in I_{N\rightarrow R}$.
    \item There exists $j < k+1$ such that $A_j = \frwd{N}{R}{r}$. Then by the inductive hypothesis we have that $r \in I_{N\rightarrow R}$.
\end{enumerate}

\noindent
Case $A_{k+1} = \slct{R}{r}$, so we have to show that $r \in I_R$. From the trace validity axioms, we know that there exists $j < k+1, r' \in \routes$, and $N \in \routers \cup \externals$ such that $A_j = \recv{R}{N}{r'}$ and $r = \import(N \rightarrow R, r')$. From the inductive hypothesis, we know that $r' \in I_{N\rightarrow R}$. Therefore by the Import check in $C$ for $N \rightarrow R$, we can conclude that  $r \in I_R$.

\noindent
Case $A_{k+1} = \frwd{R}{N}{r}$, so we have to show that $r \in I_R$. By the trace validity axioms, either:
\begin{enumerate}
    \item $r \in \originate(R \rightarrow N)$. Then from the Originate check in $C$ for $R \rightarrow N$ we have that $r \in I_{R \rightarrow N}$.
    \item There exists $j < k+1$ and  $r' \in \routes$ such that $A_j = \slct{R}{r'}$ and $r = \export(R \rightarrow N, r')$. From the inductive hypothesis, we have that $r' \in I_R$. Then from the Export check in $C$ for $R \rightarrow N$, we can conclude that $r\in I_{R \rightarrow N}$.
\end{enumerate}

Now we prove the correctness theorem for \tool, which says that \tool's checks are sufficient to ensure that the given network property holds, for all valid traces.

\noindent
\textbf{Theorem:} Given a BGP topology and policy, a network property $(\ell, P)$, and network invariants $I$, let $C$ be the set of Import, Export, and Originate checks that \tool\ generates.  
If all checks in $C$ pass and $I_\ell \subseteq P$, then for all $T \in \valid, r \in \routes, R, N \in \routers$:
\begin{itemize}
    \item if $\ell = R$ and $\slct{R}{r} \in T$, then $r \in P$
    \item if $\ell = R \rightarrow N$ and $\frwd{R}{N}{r} \in T \lor \recv{N}{R}{r}\in T$, then $r \in P$
\end{itemize}

\noindent
\textbf{Proof:} There are two cases:
\begin{enumerate}
    \item $\ell = R$ and $\slct{R}{r} \in T$. From the earlier lemma we have that $r \in I_\ell$, and since $I_\ell \subseteq P$ it follows that $r \in P$.
    \item $\ell = R \rightarrow N$ and $\frwd{R}{N}{r} \in T \lor \recv{N}{R}{r}\in T$. Again from the earlier lemma we have that $r \in I_\ell$, and since $I_\ell \subseteq P$ it follows that $r \in P$.
\end{enumerate}

Note that our reasoning does not depend on BGP converging as traces can be infinite. 
\section{Correctness Proof for Liveness}\label{sec:liveness-proof}

In this section, we prove the correctness of the modular checks for liveness properties. 

\noindent
\textbf{Theorem:} Given the following:
\begin{itemize}
    \item The network configurations
    \item A liveness property $(\ell, P)$
    \item A path $S = (\ell_1, \ell_2, \dots, \ell_{n-1}, \ell_n = \ell)$
    \item A constraint for each location $C_1 \dots C_n$
\end{itemize}
For all valid traces $T$, if all of the following are true:
\begin{enumerate}
    \item all checks (propagation, no interference) pass
    \item there exists $r$ such that $\texttt{recv}(\ell_1, r) \in T \land C_1(r)$
    \item for all $r$, $C_n(r) \implies P(r)$
    \item there are no link failures along the path
\end{enumerate}
then there exists $r'$ such that either:
\begin{itemize}
    \item $\ell \in \routers$ and there exists $r'$ such that $\slct{\ell}{r'} \in T$ and $P(r')$ holds, or
    \item $\ell \in \edges$ and there exists $r'$ such that $\texttt{frwd}(\ell, r') \in T$ and $P(r')$ holds
\end{itemize} 
\noindent
\textbf{Proof:}
Consider a valid trace $T$. By the assumption, there exists $r_1$ such that $\texttt{recv}(\ell_1, r_1) \in T$ and $C_1(r_1)$

There must exists at least one router $R = \ell_j$ and a route $r_j$ such that $\slct{R}{r_j}$ is in the trace and $\texttt{Prefix}(r_j) = \texttt{Prefix}(r_1)$. If there are no routers outside the path that have their routes accepted then $r_2 = \import(\ell_1, r_1)$ is the most prefered route at $\ell_2$, so $\slct{\ell_2}{r_2}$ will be in the trace. If there are routers outside the path that have their routes accepted, then by the no interference check, it must be that the router accepted at $\ell_j$ will satisfy $C_j(r_j)$.

Consider the last router that accepts a route from a neighbor outside the path. We will use induction to show that all locations $\ell_i$ between it and the end will have a route satisfying $C_i$:

\noindent
\textbf{Base case:} Take the last router $R = \ell_j$, where there exists $r_j, C_j$ such that $\texttt{slct}(\ell_j, r_j) \in T$ and $C_1(r_j)$. We have shown above that there must be one.

\noindent
\textbf{Inductive step:} If $\ell_i = R \in \routers$, then we know that the event $\slct{\ell_i}{r_i} \in T$ and $C_i(r_i)$ from the inductive hypothesis. We want to show that there exists $r_{i+1}$ such that $\texttt{frwd}(\ell_{i+1}, r_{i+1}) \in T$ and $C_{i+1}(r_{i+1})$.
This is true because:
\begin{itemize}
    \item let $r' = \export(\ell_{i+1}, r_i)$
    \item $\slct{\ell_i}{r_i} \in T$ and $C_i(r_i)$ (from the inductive hypothesis)
    \item $r' \neq \reject$ and $C_{i+1}(r_{i+1})$ (from the propagation check)
    \item $\texttt{frwd}(\ell_{i+1}, r_{i+1}) \in T$ (from the liveness axiom)
\end{itemize}
If $l_i = N \rightarrow R \in \edges$, then we know that $\texttt{frwd}(\ell_{i}, r_{i}) \in T$ and $C_i(r_i)$, and we want to show that there exists $r_{i+1}$ such that $\slct{\ell_{i+1}}{r_{i+1}} \in T$ and $C_{i+1}(r_{i+1})$. This holds because:

\begin{itemize}
    \item let $r_{i+1} = \import(\ell_{i}, r_i)$
    \item $\texttt{recv}(\ell_{i}, r_{i}) \in T$ (from liveness axiom given no link failures) 
    \item $r_{i+1} \neq \reject$ and $C_{i+1}(r_{i+1})$ (from the propagation check)
    \item We know that R and any router after R in the path did not accept any routes from any neighbors not in the path, so $N\rightarrow R$, so we know $\slct{\ell_{i+1}}{r_{i+1}} \in T$ and $C_{i+1}(r')$
\end{itemize}

From this, we know that at $\ell_n$, there exists a route $r_n$ such that $C_n(r_n)$ and either $\texttt{frwd}(\ell_{n}, r_{n}) \in T$ or $\slct{\ell_n}{r_n} \in T$. $C_n(r_n) \implies P(r_n)$, which is what we wanted to prove. Again, note that our reasoning does not depend on BGP converging as traces can be infinite.
\\

\end{document}